\documentclass[sigconf]{acmart}

\usepackage{booktabs} 
\usepackage{enumitem}
\usepackage{qtree}
\usepackage{tikz}
\usepackage{pgfplots}
\usepackage{graphicx}
\usetikzlibrary{arrows,shapes,positioning,shadows,trees}
\usetikzlibrary{intersections}
\usepgfplotslibrary{fillbetween}
\pgfplotsset{compat=1.17}

\setcopyright{acmcopyright}

\acmDOI{xx.xxx/xxx_x}

\acmISBN{978-1-4503-8104-8/21/03}

\acmConference[SAC'21]{ACM SAC Conference}{March 22-March 26, 2021}{Gwangju, South Korea}
\acmYear{2021}
\copyrightyear{2021}

\acmArticle{4}
\acmPrice{15.00}

\begin{document}
\title{Microeconomic Foundations of Decentralised Organisations}

\author{Mauricio Jacobo-Romero$^1$, Andr\'e Freitas$^{1,2}$}
\affiliation{%
  \institution{${}^1$The University of Manchester}
  \streetaddress{Oxford Rd}
  \city{Manchester} 
  \country{UK}
  \postcode{M13 9PL}
}
\email{{mauricio.jacoboromero,  andre.freitas}@manchester.ac.uk}
\affiliation{%
  \institution{${}^2$Idiap Research Institute}
  \streetaddress{Oxford Rd}
  \city{Martigny} 
  \country{Switzerland}
}
\email{andre.freitas@idiap.ch}

\renewcommand{\shortauthors}{M. Jacobo-Romero \& A. Freitas}

\begin{abstract}
In this article, we analyse how decentralised digital infrastructures can provide a fundamental change in the structure and dynamics of organisations. The works of R.H. Coase and M. Olson, on the nature of the firm and the logic of collective action, respectively, are revisited under the light of these emerging new digital foundations. This study proposes a novel analysis of how these instruments can affect the fundamental assumption on the role of organisations, either private or public, as a mechanism for the coordination of labour. We propose that these technologies can fundamentally affect: (i) the distribution of rewards within an organisation and (ii) the structure of its transaction costs. These changes bring the potential for addressing some of the trade-offs between the private and public sectors.
\end{abstract}

%
%



\maketitle

\section{Introduction}

 The purpose of this paper is to analyse how distributed digital infrastructures will effect structural changes in the nature of organisations. We aim to qualify the microeconomic foundations of decentralised organisations formally. This analysis is grounded in the works of Coase\cite{Coase1937} and Olson\cite{Olson1965}. Both works consolidate into a set of organisation roles, and relations that define a \textit{firm} and support its existence.  In one hand, Coase proposed a set of necessary attributes and conditions for the \textit{firm} to exist and operate. On the other, Olson modelled the economics of coordinated group actions. These works provide two complementary economic reference frames which jointly answers why centralised private companies are the prevalent model of economic organisation.
 
 By putting these two perspectives to dialogue, we derive a mathematical model which elicits the cost and reward structures involved in the coordination of the different agents behind the firm. This model is then used as a foundation to understand how distributed digital infrastructures (DDIs) can disrupt this dominant centralised model. 
 
 We particularly focus on the construction of the microeconomic argument that two fundamental components for enabling fully decentralised organisations are \textit{smart contracts} and a \textit{reward distribution metric}. We argue that these two types of digital infrastructure can promote a bottom-up model of organisation.
 
 Two corollaries are derived: (i) the emulation of the role of the entrepreneur and investor by a distributed network of agents and (ii) the emergence of a new type of organisation (distributed firms) which merges attributes from the private and public sectors.
 
 This work is organised as follows: the second, third and fourth parts describe our mathematical interpretation of Coase's and Olson's work, respectively. The fifth section contains our operation framework. The sixth, seventh and eighth components introduce the analysis, discussion and a numerical example respectively. Lastly, we present our conclusions.
 
\section{Background}  
\subsection{Distributed ledgers}  

 In order to record transactions, firms employ the double-entry bookkeeping system. The basic unit of this method is the account. An account organises transactions according to a rule called T-shape. All these transactions are, then, kept in a memoir known as {\it ledger} \cite{Atrill2017}. Ledgers were introduced in the 15\textsuperscript{th} century by the Venetian Republic \cite{Davidson2016} as a centralised information solution.

 Despite the benefits of decentralisation, ledgers kept the same operation principle even with computer automation. It was until the introduction of {\it digital currencies} that distributed ledger technology became observed. A Distributed Ledger, therefore, is a shared database that stores assets of multiple sites, geographies or institutions \cite{UKGovernmentChiefScientificAdviser2016}. 

 Blockchain systems integrate a distributed ledger implementation as part of their protocol. Thus, it is possible to eliminate the need for trust and promote effective teamwork and cooperative activities \cite{Davidson2016}.

\subsection{Blockchain economic models }  
 The introduction of blockchain technology has underpinned the evolution of productive mechanisms in the form of new organisation types \cite{Pazaitis2017}. According to Mary S. Morgan, an economic model is a theoretical approach to describe the operation of productive processes \cite{Morgan2008}. Therefore, Blockchain has enabled the launch of new economic models.

 These new structures diminish the problem of contractual incompleteness considerably \cite{Pazaitis2017}. Blockchain systems introduce "a spontaneous order produced by people within the rules of the law of property, tort and contract with not a particular end", or Catallaxy \cite{Davidson2016}. A Catallaxy is composed of multiple agents within an "extended order". These agents are social and governed by social rules. They own specialised knowledge, shape their plans, and are mutually coordinated through the market price system under scarcity environments \cite{Hayek1988} \cite{Buchanan1990}.

 Some examples of this new structure are Backfeed and La `Zooz. Backfeed promotes a rewarding system protocol, for any industry, based upon economic tokens and reputation scores. This mechanism is called Proof-of-Value \cite{BACKFEED}. They construct complex light structures that do not rely on classes but individual agents who are self-coordinated \cite{Marsh2008}. In the case of La `Zooz, the company offers mining services based upon a token model. La `Zooz aims to utilise the empty seats in regular car users. Users receive tokens when they integrate new members into the community and when they share their car with other users\cite{SOKOLOWSKY2015}.

 Both companies have minimised the cost of legal services, explicitly speaking contract cost, by allowing their members to establish "organic" coordination over the Blockchain with the minimum number of legal agreements.  In this fashion,  both organisations have empirically integrated the principles of the "nature of the firm "\cite{Coase1937}.

 \section{Summary of the Proposed Model}  
 This model pioneers on whether \textit{emerging digital infrastructures can heighten bottom-up economic organisation in highly digital settings, through minimising the free-rider problem, as posed by Olson and emulating the microeconomic structure described by Coase}. We argue that digital infrastructures facilitate the creation of a large-scale digital cooperative model by changing the microeconomic dynamics in the organisation structure. The following is a summary of the key claims that underpin this exposition:
 
 \begin{itemize}
        \item \textit{Smart contracts} can minimise the difference between internal and external transaction costs within the firm by facilitating the creation of distributed production chains, and opportunistic and fragmented labour relationships. As a result, the need for a fixed-cost structure of labour is minimum. This notion changes the fundamental assumption behind Coase's justification for the existence of the firm.
       
        \item \textit{Distributed apps (DAPPs) associated with productivity/reward metrics} can provide an infrastructure that recompenses digital productivity at a fine-grained scale. DAPPs will also grant the possibility of measuring and gratifying contributions (at a task/component level). A contractual framework formalises this participation by (i) minimising the free-rider problem as posed by Olson and maximising the value to the productive individual ; (ii) reducing the need for a hierarchical managerial structure, and providing distributed accountability.
        
        \item The possibility of \textit{distributed ownership} allows the organisation to partially emulate the role of the entrepreneur, investor, and the associated managerial hierarchy. As a result,  this model brings fuzziness among these functions.
 \end{itemize}

 The core of the previous postulates lays in the following assumptions:

 \begin{itemize}
    \item Smarts contracts are enforceable.
    
    \item Reward metrics are transparent, minimally efficient and are part of the contractual agreement. That is, productivity evaluation has an acceptable accuracy level, and the contract clauses include these performance indicators. 
    
    \item Processes and tasks are specific and capable of being delegated.
 \end{itemize}
 
 The following sections outline our mathematical interpretation of Coase's and Olson's work, and a business assessment model. Afterwards, we explain the proposed framework.

\section{Coase's Nature of the Firm}  
\subsection{Outline}

 Coase's fundamental contribution, to the nature of the firm, is the idea that within a company a force, under the figure of an agent, partially neutralise the price system and the market forces.  

 The role of the "entrepreneur" is responsible for creating this "environment". In this fashion, the "entrepreneur" should generate the conditions that enable price-system independent economic transactions among different company's entities \cite{Suematsu2014}. Thus, a pool of human and material resources, within the firm, is available for allocation and free from the market or external \textit{transaction costs} - e.g. freight price and contract negotiation.
 
  Therefore,  the transaction cost is an essential concept of Coase's study \cite{Coase1937}.
 
\subsubsection{Formalisation - Minimising Transaction Costs} ~\\
 After reviewing Coase's work \cite{Coase1937}, we considered that the following ideas are the most relevant to our research.
 
 \begin{enumerate}[label=\roman*)]
        \item {\it "But in view of the fact that it is usually argued that co-ordination will be done by the price mechanism, why is such organisation necessary? Why are there these "islands of conscious power"? Outside the firm, price movements direct production, which is co-ordinated through a series of exchange transactions on the market. Within a firm, these market transactions are eliminated and in place of the complicated market structure with exchange transactions is substituted the entrepreneur-co-ordinator, who directs production"} \cite{Coase1937}. Thus,
 
          \begin{enumerate}
          
              \item In the Market:
                    \begin{equation}
                           \phi = f(P, ETC, t) 
                    \end{equation}
                    Where $\phi$ represents production, $P$ is the price, $ETC$ stands for external transaction cost, and $t$ is time. 
                    
              \item In the firm:
                    \begin{equation}
                           \phi = f(TTC, t) 
                    \end{equation}
                    Where $\phi$ describes production, $TTC$ represents the total transaction cost, and $t$ is time.
          \end{enumerate}
          
        \item {\it "Firms might also exist if purchasers preferred commodities which are produced by firms to those not so produced; but even in spheres where one would expect such preferences (if they exist) to be of negligible importance, firms are to be found in the real world. Therefore there must be other elements involved. The main reason why it is profitable to establish a firm would seem to be that there is a cost of using the price mechanism. The most obvious cost of "organising" production through the price mechanism is that of discovering what the relevant prices are. This cost may be reduced but it will not be eliminated by the emergence of specialists who will sell this information"} \cite{Coase1937}. Then,
          \begin{equation}
                ETC > ITC
          \end{equation}
          \begin{equation}
                TTC = ETC + ITC
                \label{TTC}
          \end{equation}
          Where ETC stands for External Transaction Cost, ITC is the Internal Transaction Cost, and TTC is the Total Transaction Cost.
          
        \item {\it "The operation of a market costs something and by forming an organisation and allowing some authority (an "entrepreneur") to direct the resources, certain marketing costs are saved. The entrepreneur has to carry out his function at less cost, taking into account the fact that he may get factors of production at a lower price than the market transactions which he supersedes, because it is always possible to revert to the open market if he fails to do this"} \cite{Coase1937}. Hence,
          \begin{equation}
                    F_{M} = \Big\{ \exists f(ETC, ITC, U_e)  \mathrel{\Big|}  ETC \geq ITC \Big\}
          \end{equation}
          Where $F_{M}$ represent the firm and $U_e$ stands for the firm's operation uncertainty.

       \item {\it "One entrepreneur may sell his services to another for a certain sum of money, while the payment to his employees may be mainly or wholly a share in profits"} \cite{Coase1937}. Therefore,
          \begin{equation}
                C_l =  \sum_{i=1}^{n} W_i = \Delta \pi
                \label{COST_OF_LABOR}
          \end{equation}
          Where $C_l$ represents the cost of labor, $W_i$ is the wage of each individual, $n$ equals the number of members/employees of the group, and $\Delta \pi$ is a portion of the profit.
         
       \item {\it "With uncertainty entirely absent, every individual being in possession of perfect knowledge of the situation, there would be no occasion for anything of the nature of responsible management or control of productive activity. Even marketing transactions in any realistic sense would not be found. The flow of raw materials and productive services to the consumer would be entirely automatic"} \cite{Coase1937}. Consequently,
          \begin{equation}
                F_{M} = \Big\{ \exists f(ETC, ITC, U_e)  \mathrel{\Big|} U_e \neq 0 \Big\}
          \end{equation}
          Where $F_{M}$ represent the firm and $U_e$ stands for the firm's operation uncertainty.

       \item {\it "a firm will tend to expand until the costs of organising an extra transaction within the firm become equal to the costs of carrying out the same transaction by means of an exchange on the open market or the costs of organising in another firm"} \cite{Coase1937}. Then,
          \begin{equation}
              F_{M} > 0 \thickspace  \Leftrightarrow \thickspace MITC \leq METC
          \end{equation}
          Where MITC is the Marginal Internal Transaction Costs, and METC stands for Marginal External Transaction Costs.
 \end{enumerate}
 
 From these ideas, we have identified the following fundamental attributes of the firm:
       \begin{enumerate}[label=\textbf{A\arabic*.}]
            \item The presence of a $ITC$ and uncertainty $U_e$ are necessary to the existence of a firm.
            \item The firm emerges and expands as a result of the transaction cost reduction.
            \item \label{structure} The firm is composed of a system of relationships which are under the authority of an entrepreneur \cite{Coase1937}. This agent is, then,  responsible for coordinating the firm's infrastructure to: 
                \begin{enumerate}[label=\roman*)]
                        \item measure and estimate the product price, the internal/external costs at a time $t$, and 
                        \item intervene in the functional-cost composition of the firm via \textit{productivity expansion/contraction}, and \textit{transaction selectivity}.
                \end{enumerate}
            \item The size of the firm leads to capital accumulation.
            \item The hierarchy induced by \ref{structure} encodes the asymmetry of information associated with the price-costs. 
            \item Besides information asymmetry, a complex hierarchical structure fosters the specialisation of the cost structure. This specialisation is the result of knowledge accumulation.
       \end{enumerate}

\section{Olson's Logic of Collective Action}
\subsection{Outline}
 Olson states that the {\it"value"} to the individuals governs the dynamics of a workgroup \cite{Olson1965}. Hence, the firm's financial and operational behaviour is closely related to this variable. He also observed that different groups exhibit distinct "benefits" for their members. This phenomenon drove Olson to classify groups, and firms, according to their size, in terms of value, into small, and large/latent \cite{Olson1965}.
 
 \textit{Small groups} offer considerable value to individuals \cite{Olson1965}. On the other hand, the \textit{latent groups} reward small benefits to their members since their interest are notably fragmented; members tend to work motivated by their objectives instead of the common benefit. Therefore, small groups are more efficient than large ones. However, the small groups develop monopolistic practices and a biased wealth distribution \cite{Olson1965}. As for latent groups, they manifest a lack of efficiency and a remarkable difficulty to adopt changes\cite{Olson1965}. 

 With this theoretical background, it is possible to observe the business environment and identify groups and their singularities.
 \subsection{Formalisation: The Free Rider Problem}
 Olson developed a tight mathematical explanation on small group behaviours. Though, he did not provide an analytical narrative for latent groups. Therefore, we revisited his analysis and proposed the following interpretation:

\begin{enumerate}[label=\roman*)]
        \item {\it "The larger the group, the smaller the fraction of the total group benefit any person acting in the group interest receives, and the less adequate the reward for any group-oriented action, and the farther the group falls short of getting an optimal supply of the collective good, even if it should get some"} \cite{Olson1965}. 
        \begin{align*}
            \begin{split}
                S_g \propto \frac{1}{F_i} \:,
           \:   S_g \propto \frac{1}{V_i} \:, \:   \text{and} \:
           \:   S_g \propto \frac{1}{\zeta_o} .
            \end{split}
         \end{align*}
        If we represent these proportionality relations as equations, we can reformulate these expressions as:
        \begin{equation}
             S_g = k_o \frac{V_g}{V_i}  ,
             \label{EQ_KO}
         \end{equation}
         \begin{equation}
             S_g = k_g \frac{1}{V_i}  ,
             \label{EQ_KG}
         \end{equation}
         and
         \begin{equation}
             S_g = k_s \frac{1}{\zeta_o}. 
         \end{equation}
 \noindent where $S_g$ represents the 'size' of the group, $F_i$ denotes the gain that an individual might get from the group earnings, $\zeta_o$ is the service supply at the equilibrium point, $k_o$ and $k_g$ are constants which compute the order of the group, and $k_s$ is a constant that determines the order the group at the optimum service supply point. Note that ($F_i$) equals $\frac{V_i}{V_g}$ where $V_i$ is the value to the member, individual, and it is equal to $F_i S_g T$. $T$ is the rate at which the collective good is obtained \cite{Olson1965}. If we analyze equations \ref{EQ_KO} and \ref{EQ_KG}, we found they have a similar structure; thus, we can represent $k_g$ in terms of $k_o$:
          \begin{equation}
               k_g = k_o V_g
               \label{DEF_KG}
          \end{equation}
 From \ref{DEF_KG}, we can infer that $k_o$ should take a value between zero and one, $0 < k_o \leq 1$, since there are no other elements that might amplify the value of $V_g$. Hence, we can assert that a value of zero means that no member of the group is pursuing the common goal. Whereas, a value of one indicates that the objective is alluring to every member of the team. Thus, equation \ref{EQ_KG} is a particular case of \ref{EQ_KO}. 
         
 \item {\it "The larger the group, the smaller the share of the total benefit going to any individual, or rto any (absolutely) small subset of members of the group, much less any single individual, will gain enough from getting the collective good to bear the burden of providing even a small amount of it; in other words, the larger the group the smaller the likelihood of oligopolistic interaction that might help to obtain the good"} \cite{Olson1965}. Therefore:
        \begin{equation*}
             S_g \propto \frac{1}{P_o} 
        \end{equation*}
         Thus,
        \begin{equation}
             S_g = \frac{k_v}{P_o} 
        \end{equation}
 \noindent where $P_o$ stands for the probability of oligopolistic interaction to provide the service and $k_v$ is a constant of proportionality that represents the potential group size. 
        
         \item {\it "The larger the number of members in the group the greater the organisation costs and thus the higher the hurdle that must be jumped before any of the collective good at all can be obtained. For these reasons, the larger the group the farther it will fall short of providing an optimal supply of a collective good, and very large groups normally will not, in the absence of coercion or separate, outside incentives, provide themselves with even minimal amounts of collective good." } \cite{Olson1965}. In this case:
         \begin{equation*}
             S_g \propto C_o 
         \end{equation*}
        Repeating the same methodology:
        \begin{equation}
             S_g = k_{\omega} C_o 
             \label{KOMEGA}
        \end{equation}
 \noindent where $C_o$ represents the cost of organisation, and $k_{\omega}$ is constant of proportionality. From equation \ref{KOMEGA}, we can see that optimum values for  $k_{\omega}$ are $k_{\omega} > 1$. When $k_{\omega}$ gets a value from the interval: $0 < k_{\omega} \leq 1$, we can consider that the company is underperforming.
        
        \item {\it "if there is one person in a group who is willing to bear the entire costs of providing a service, the action is likely to be executed or the service provided"} \cite{Olson1965}.
           \begin{equation}
               V_i > C  \implies \zeta > 0
               \label{VIANDCOST}
           \end{equation}
        \item On the other hand, if there is a small number of enthusiastic persons prepared to bear the full costs of providing a collective good, this may not be provided \cite{Olson1965}.
           \begin{equation}
               P(V_g^*) \to 0 \text{,}  \quad \textrm{if n} < \textrm{m}
           \end{equation}
        where $V_g^*$ is the value of the group with the new service, $P(V_g^*)$ is the probability that the new service would be provided to the group, n is the number of enthusiastic persons and m the total number of group members.
        
        \end{enumerate} 
        
 Previous statements uncover the following:
 
            \begin{enumerate}[label=\textbf{B\arabic*.}]
                \item Equation \ref{VIANDCOST} describes the required conditions for the free-rider problem to arise. The free-rider problem occurs when a member of a group uses or obtains profits from a group benefit such as public goods or services with no payment. This misapplication may result in an underprovision, or even cancellation, of those profits.
                \item A group member will be committed to organisation objectives only if he can obtain benefits from his effort on supplying current or new services.
                \item Organisation performance can be improved if group members obtain their expected benefits.
                \item Group members benefits depend upon personal preferences.
                \item Each constant of proportionality, $k$, represents different concepts. In other words, they describe several phenomena; their units reveal this:
                \begin{enumerate}[label=\roman*)]
                    \item $k_o$ - $\pounds$,
                    \item $k_g$ - $\pounds^2$,
                    \item $k_s$ - $u\pounds$,
                    \item $k_v$ - $\pounds$, and
                    \item $k_w$ - it is a dimensionless quantity.
                \end{enumerate}
                where $u$ stands for product units and $\pounds$ is monetary value.
          \end{enumerate}

 \section{Firm's performance}

\begin{center}
        \begin{figure}
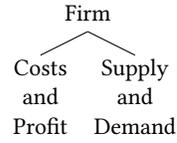

            \centering{
                   \Tree[.Firm {Costs\\and\\Profit} {Supply\\and\\Demand} ]} 
       \caption{Firm's business performance variables.}
        \label{label1}
        \end{figure}
    \end{center}

 There is no consensus on how to measure business performance \cite{Atrill2017}\cite{Team2017}. For our analysis, we explored the branches presented in Fig.  \ref{label1}. From this chart, we can observe that different variables can describe the firm's financial performance. In the next subsections, we will explain our development of Coase's and Olson's work in these terms.

\subsection{ Costs and profit }
 According to the classic economic theory, the price mechanism drives the market \cite{Coase1937}. Thus, it is plausible to state that price, $P$, is the primary variable of this system. According to Coase's remarks, we can describe price, then, as the sum of transaction costs for a specific product or service:
    \begin{equation}
        P = \sum_{i=1}^{n} t_{c_i} 
    \end{equation}
 \noindent where ${t_{c_i}}$ is the transaction cost. We can extend this definition making $t_c$ equals to $TTC$ from \ref{TTC}. Thus,
    \begin{equation}
        P = \sum_{i=1}^{n} TTC_{i} = \sum_{i=1}^{n} \big (ETC_i + ITC_i \big )
        \label{PRICE_EQ}
    \end{equation}
 It is relevant to mention that $ITC_i$ will be equal to 0 when describing prices in the market system.
    \begin{definition}
         $ETC_i$ is a composition of charges in land, labour, and capital. That is:
            \begin{equation}
                 ETC_{i} = ( 1 + U_p ) [  C_{land_i} + C_{labour_i} + C_{capital_i} ] 
            \end{equation}
    \end{definition}
 
 \textit{Remark.} We considered that price uncertainty, $U_{p}$, equals the expected inflation of the market during the evaluated period for the transaction $i$.

 \begin{definition}
 As for $ITC_i$:  
    \begin{equation}
      ITC_{i} = l_{c_i} + C_{l_i} + U_{O_i}
      \label{ITC_I_DEF}
    \end{equation}
 \noindent where $l_{c_i}$ is the legal cost, $C_{l_i}$ is the organisation cost, and $U_{O_i}$ operational uncertainty for the transaction $i$.
 \end{definition} 
 Legal cost, $l_{c_i}$, refers to the cost of collectively securing the firm's privately defined objectives. In other words, the cost associated with protecting the private information that is not efficiently secured through simple market exchanges. The value of these agreements will depend upon both frequency and the nature of the trading organisation. The longer the contract, the lower the cost \cite{Williamson1998}. 
 
 On the other hand, $C_{l_i}$ is defined by equation \ref{COST_OF_LABOR}. Concerning $U_O$, it can be calculated with the Hurwicz' Index (HI) \cite{Kantarelis2017}. This indicator is employed when the firm is not either 100\% optimistic or pessimistic. HI definition is then:
    \begin{equation}
        HI = m + s = 1
    \end{equation}
 \noindent where $m$ is the degree of optimism and $s$ is the degree of pessimism. HI is applied to the expected benefits of the produced product. Thus, the option with the best HI index value is selected and then applied to the $ITC$.  

 \subsection{  Supply and demand }
 Employing the classical microeconomics {\it "supply and demand"} mathematical model, we have the following:
    \begin{equation}
        S = a + bP
    \end{equation}
    \begin{equation}
        D = c - dP + eIE
    \end{equation} 
 \noindent where $P$ is the price, $S$ is production, $D$ is demand, $IE$ represents the inflationary expectations, and $a,b,c,d$ and $e$ are {\it behavioral} variables that reflect the reaction of producers (supply) and buyers (demand) to price. 
 
\begin{figure}[tpb]
        \centering{\includegraphics[width=\linewidth]{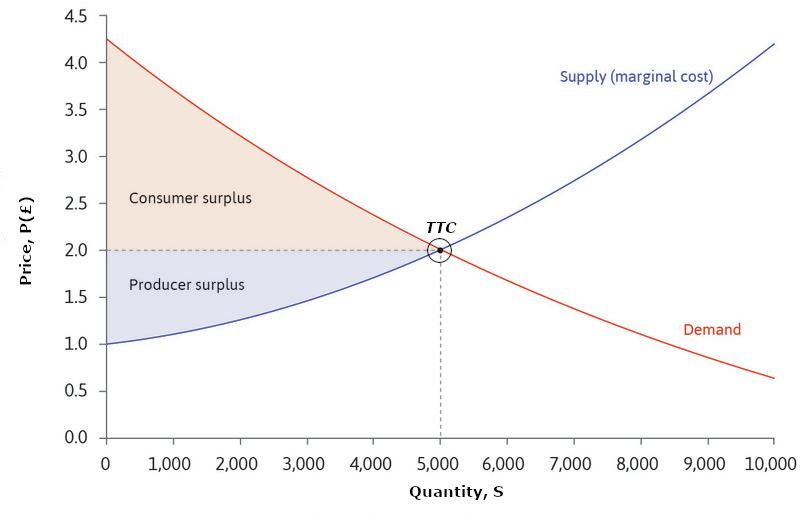}}
        \caption{TTC in Production (S) and Demand (D) terms.}
        \label{label_TTC_SD}
    \end{figure}

 At the market equilibrium point, $S$ and $D$ are equal. If we also substitute $IE$ for $U_p$, since they represent the same variable, then we have that price can be expressed by:
    \begin{equation}
        P = \frac{c+eU_p-a}{b+d}
        \label{PRICE_PD}
    \end{equation}
 \noindent From equation \ref{PRICE_EQ}, we can rewrite equation \ref{PRICE_PD} as follows:
    \begin{equation}
        \sum_{i=1}^{n} \big (ETC_i \big ) + \sum_{i=1}^{n} \big ( ITC_i \big ) = \frac{c+eU_p-a}{b+d}
        \label{TTC_EQ_PRICE}
    \end{equation}
 \noindent From \ref{TTC_EQ_PRICE}, we can see that $TTC$ is not affected by Production and Demand themselves, but by market conditions which are represented by the behavioural constants. See figure \ref{label_TTC_SD}.
     \begin{figure*}[tpb]
        \centering{\includegraphics[scale=0.5]{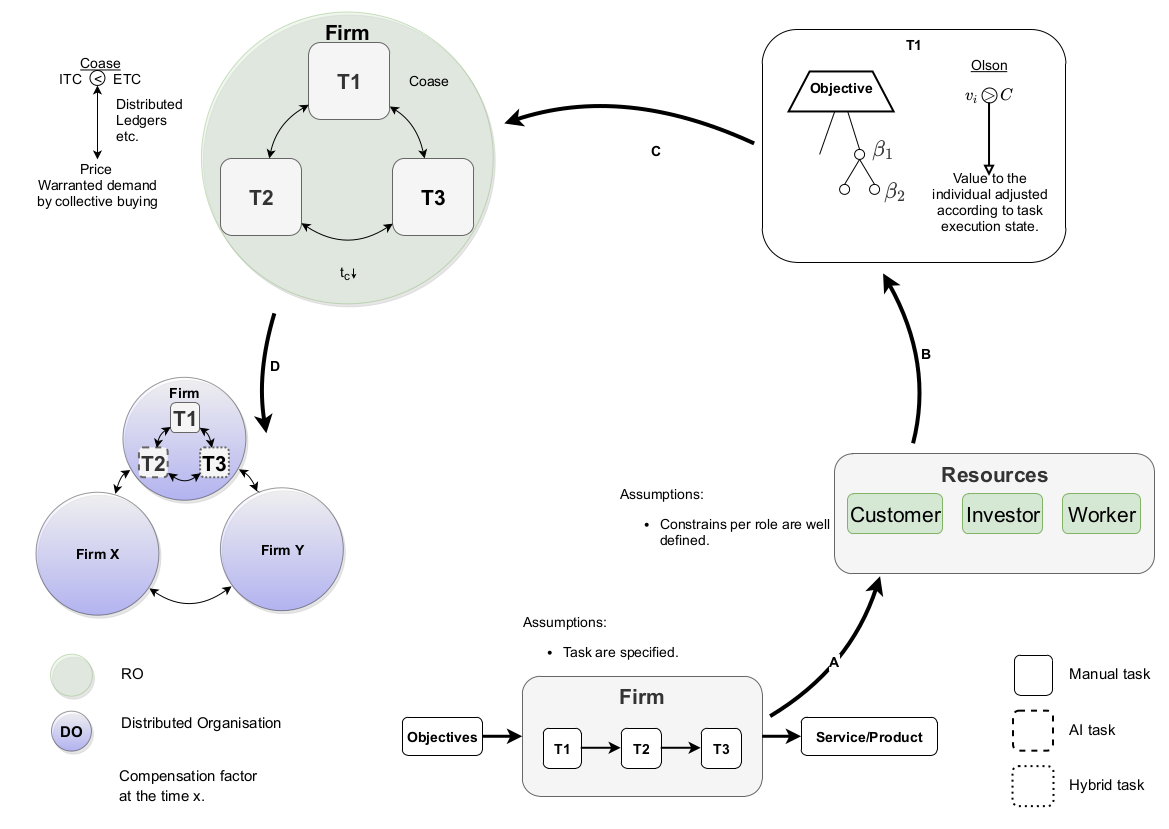}}
        \caption{Model information flow.}
        \label{label_AIMs}
     \end{figure*}

 
%
 \section{ Proposed Framework }

 This work introduces a model for a digital cooperative. We define a digital cooperative as a decentralised organisation that aims for balancing the distribution of the rewards and values to the individuals of the group. At the same time, the proposed infrastructure aims to minimise the transaction costs with the support of smart contracts.
 
 The {\it "Distributed Organisation"} (DO) is the abstract idealisation behind the digital cooperative. This model satisfies Coase's conditions to exist and, at the same time, monitors and adjusts the value of the individuals within the new firm in order to minimise Olson's free-rider problem. See figure \ref{label_AIMs}. 
 
 The DO framework consists of four main stages:
     \begin{enumerate}[label=\Alph*)]
        \item Task formalisation and estimation,
        \item Resource allocation,
        \item Olson's condition implementation,
        \item Coase's prerequisite monitoring, and
        \item Distributed infrastructure implementation.
    \end{enumerate} 
 \subsection{Task formalisation and estimation, and resource allocation} 

 We start with the assumption that a digital firm is a composition of hierarchical business processes distributed as human-mediated practices and digital artefacts. Besides, our model assumes that tasks can be formalised and estimated within a certain margin of error. That is, duties are decomposable and described by formal specifications.
   
 The second stage assigns the resources for the development of the business process components. Three roles are associated with the materialisation of these parts: worker, investor and customer. The component-level assignment and accountability associated with smart contracts reduce the barriers for agents to engage with more granular processes and components bringing ambiguity across roles. Workers can be rewarded via a wage relationship, associated with the revenue-cost structure of the firm, or engaged as a voluntary worker; i.e. in an investor capacity. Similarly, customers can join as investors in a crowd-funding type of relation. DO integrates role-budget constraints in the following manner:
    \begin{itemize}
       \item Investor:
            \begin{equation}
                M_i \leq \sum_{i=1}^{n} TTC_i
            \end{equation}
            \noindent where $M_i$ is the investor budget.
       \item Customer:
            \begin{align}
                \begin{split}
                    M_c \leq P  \\
                    M_c \leq \frac{c+eU_p-a}{b+d}
                \end{split}
            \end{align}
            \noindent where $M_c$ is the customer budget.
       \item Worker:
            \begin{equation}
                 M_m \geq w_m
            \end{equation}
            \noindent where $M_m$ is the customer budget and $w_m$ her wage. 
       \end{itemize}
 \subsection{Olson's conditions}
 In order to balance the benefits of each group member, we extended the description of $V_i$. In this fashion, we define the variable that $E$ will adopt the following values:
        \[
            E =  \begin{cases}
                        0 &\quad\text{ if } E \text{ is an investor,} \\
                        1 &\quad\text{ if } E \text{ is an employee.}
                 \end{cases}
         \]
 We also assumed that an employee depends only on his wage to cover his basic needs. Thus, they would receive their wage plus dividends. Whereas in the case of the investors, they would only receive the dividends. Thus, dividends will help the Reward Oracle (RO) to balance the {\it "new firm"} organisation dynamics. 
 
 Hence, we can express $V_{Ej}$ as follows: 
         \begin{equation}
            V_{Ej} = W_{Ej} +  \big[ \beta_{Ej} (S - C ) \big]
            \label{COASE_OLSON_EQU2}
         \end{equation}  
 Where $W_{Ej}$ is the wage of the individual, $\beta_{Ej}$ stands for the dividends of the individual, $S$ is the total sales, and $C$ is the total costs of the firm. $W_{Ej}$, then, behaves in the following manner: 
         \[ 
            W_{Ej} = 
                \begin{cases}
                    W_{0j} = 0, &\quad\text{ if } E = 0 \text{,}\\
                    W_{1j} \neq 0,  &\quad\text{ if } E = 1 \text{.}
                \end{cases}
        \]
 If we analyse $W_{Ej}$ under the game theory standpoint, we can use the following optimised function \cite{Team2017}:
        \begin{equation}
            W_{1j} = \frac{w_{r_{1j}}}{1 - \alpha} 
        \end{equation}  
 Where $w_{r_{1j}}$ is the amount the employee can earn in the price market, and $\alpha$ is a the employee's effort per hour with values between $0<\alpha<1$. As for $\beta_{Ei}$, we consider the following condition:
        \begin{equation}
            \sum\limits_{j=1}^{m} \beta_{Ej} = 1
            \label{EqBetaAdd}
        \end{equation}
 \noindent Where $m$ is the number of total persons in the firm: employees and investors.           
 If we rewrite the equation \ref{EqBetaAdd} in terms of the specific roles, we have the following: 
        \begin{equation}
            \sum_{j=1}^{m} \beta_{Ej} = 
                \begin{cases}
                    {\sum\limits_{p=1}^{a} \beta_{0p} = r}, &\quad\text{ if } E = 0 \text{,}\\
                         \\ 
                    \sum\limits_{q=1}^{b} \beta_{1q} = (1 - r),  &\quad\text{ if } E = 1 \text{.}
                \end{cases}
        \end{equation}
          
 \noindent Where $p$ is the number of total investors, $r$ stands for the return rate on the investment, and $q$ represents the number of employees. As for the allocation coefficients, $\beta$s, we considered that royalty distribution would depend on two factors. The first one, the proportion of the investment allocated to the endeavour. The second one, the contribution that a team member has done to the whole group. A transparent contractual productivity metric evaluates this contribution. The model assumes that proxies for performance can be estimated and computed in digital and distributed settings. The analysis of possible schemes to materialise performance measurement and estimation is beyond the scope of this work.
 
 Thus, we propose the following allocation functions:
        \begin{equation}
            \resizebox{\hsize}{!}{$\displaystyle{
                    \beta_{Ej} = 
                        \begin{cases}
                           \frac{rP_{0j}}{C},       &\quad\text{ if } (S - C) > 0 \quad     \text{\&} \quad E = 0 \text{,}\\
                                      \\ 
                            \Big[   \frac{\tau_{1j}}{\tau_{1{n_j}}} \Big] \Gamma(n_{1j}) (1 - r) ,  &\quad\text{ if } (S - C) > 0 \quad \text{\&} \quad E = 1 \text{,}\\
                                       \\
                                       0,  &\quad\text{ if } (S - C) \leq 0 \text{.}
                         \end{cases}
                    }
                $}
        \end{equation}
        \begin{equation}
            \resizebox{\hsize}{!}{$\displaystyle{
                \Gamma(n) = 
                    \begin{cases}
                        1, &\quad\text{ if } n = l = 1 \text{,}\\ 
                        1 - \Big[ \Phi(n-1) - \Phi(-(n-1)) \Big], &\quad\text{ if } n = l \text{,}\\
                                       \\ 
                      \Big[ \Phi(n) - \Phi(-n) \Big] - \Big[ \Phi(n-1) - \Phi(-(n-1)) \Big],  &\quad\text{ if }  1 < n < l \text{,}\\
                                       \\
                      \Big[ \Phi(n) - \Phi(-n) \Big], &\quad\text{ if } n = 1 \text{ \& } n \neq l \text{ .}
                    \end{cases}
                }
            $}
        \end{equation}
 
 \noindent Where $r$ is the rate of the promised royalty to investors, $P_{0j}$ represents the portion of the project cost granted by the investor $j$, $S$ stands for sales, $C$ expresses the costs, $\tau_{1j}$ denotes the total number of performance samples of the employee $j$, $\tau_{1{n_j}}$ embodies the full performance per organisational level, $\Phi$ is the standard normal distribution function, and $n_j$ denotes an optional hierarchical (trust) level component. If the contributor is an investor, $\beta$ calculates the interest amount this person may receive according to the share he contributed. An investor is allowed to invest up to the total cost of a project.  As for the employee,  their royalty share is calculated based on a performance metric. These values are weighted and multiplied by the normal distribution function interval. This factor aims to allocate the most significant portion of the profit to the share of the firm that incorporates more employees; in other words, we included the empirical {\it 3-sigma rule}. Lastly, if no profit is made, no one is entitled to get dividends. 

 In this fashion, the {\it RO} will be responsible for balancing equation \ref{COASE_OLSON_EQU2} and monitoring \ref{COASE_OLSON_EQU_1} to keep Coase's and Olson's conditions allowing the firm to exist and minimise the free-rider problem within a specific contextual setting. It is also worth to mention that we extended this analysis to a multiproduct/multiservice company by regarding sales and costs as marginals.
 
 Changes in both productivity and the value to the individual will help us to assess the framework. Therefore, we have the following mathematical expressions to analyse these variations:
        \begin{equation}
            P(v_i) = 1 - e^{-\alpha v_i}
        \end{equation}
        
 \noindent Where P is productivity, and $\alpha$ represents a measure of the employee fitness. $\alpha$ can take any  of the following values: 
        \begin{equation*}
            0 < \alpha < \infty   
        \end{equation*}
 A value of $0$ means that the individual does not know how to execute his duties whereas a value of $\infty$ implies that he has maximum fitness for a task set. For the value to the individual according to the organisation level, we propose the following formula:
        \begin{equation}
            v_i(l) = v_i(1) \big( \frac{r}{N} \big)l
        \end{equation}     
 \noindent Where $v_i(1)$ is the value to the individual of the lowest organisation level, $r$ is the CEO-to-worker compensation ratio, $N$ the maximum organisation levels, and $l$ is the individual organisation level. With $l > 2$.    

 \subsection{Coase's prerequisite}    
 {\it RO} integrates Coase’s requirements as follows:
        \begin{subequations}
            \begin{align}
                ITC &\leq ETC  \\
                ITC + ETC &< V_{Ej}
            \end{align}
            \label{COASE_OLSON_EQU_1}
        \end{subequations}
 ITC and ETC are marginal costs. That is, they only represent the cost of analysed services. DO will adjust the value of these variables at the end of every contractual cycle. 

 \subsection{Distributed infrastructure implementation}
 During this stage, RO classifies process activities in three groups: manual task, AI task, and Hybrid task. Task description must contain its legal and organisation costs in the form of contracts. These terms will be employed to set the conditions of the smart contracts that will ensure the minimal operation conditions for the task. DO distribution will directly impact equation \ref{ITC_I_DEF}. However, the uncertainty component is not directly affected since organisation uncertainty is an endemic characteristic of the firm. Therefore, RO's ultimate goal would be:
        \begin{equation}
            ITC_i \approx U_{O_i}
        \end{equation}
 RO will enable market communication and conditions automatically. Thus, cost and delivery conditions behaviour by the transaction will be available to the firm's management to amend any process deviation.  
 \subsection{Expected effects}
 To summarise, we expect to obtain the following results from the RO implementation:
        \begin{enumerate}[label=(\roman*)]
            \item A reduction in the free-rider problem, 
            \item A reduction in the barriers for a bottom-up organisation,
            \item The integration of the roles: consumer, investor, and employee,
            \item A decrease in capital accumulation via the distribution of the entrepreneurial functions across the company.
        \end{enumerate}

\begin{figure}[tpb]
            \centering{\includegraphics[width=\linewidth]{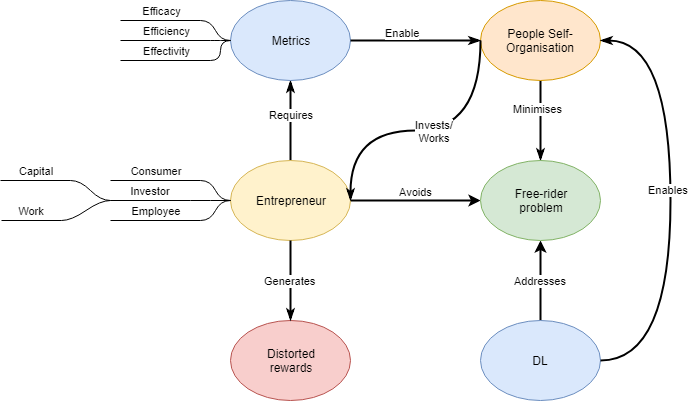}}
            \caption{Attributes of the proposed Distributed Organisation (DO).}
            \label{label_Flows}
        \end{figure}

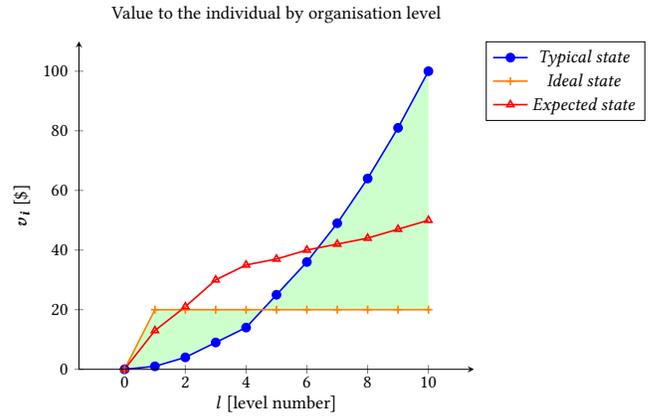
\begin{figure}[tpb]
            \resizebox{\hsize}{!}{$\displaystyle{
            \centering{
                    \begin{tikzpicture}
                        \begin{axis}[
                            title={Value to the individual by organisation level},
                            axis lines=left,
                            xlabel={$l$ [level number]},
                           ylabel={$\boldsymbol{v_i}$ [\$]},
                            enlarge x limits={abs=1.5},
                            ymin=0, ymax=110,
                            xticklabel style={
                            fill=white,
                            fill opacity=0.7,
                            text opacity=1,
                            inner sep=1pt,
                            },
                            axis on top,
                            legend pos=outer north east,
                        ]
                            \addplot [blue, thick, mark=*, name path=A] table [x={n}, y={xn}] {es.dat};
                                  \addlegendentry{\text{\it{Typical state}}}   
                            \addplot [orange, thick, mark=+, name path=B] table [x={n}, y={xn}] {ids.dat};
                              \addlegendentry{\text{\it{Ideal state}}}   
                            \addplot [red, thick,  mark=triangle, mark size=2] table [x={n}, y={xn}] {ims.dat};           \addlegendentry{\text{\it{Expected state}}}   
                            \addplot+[green, fill opacity=0.2] fill between[of=A and B,soft clip={domain=0:10}];
                        \end{axis}
                    \end{tikzpicture}
                    }
              }
        $}
        \caption{Estimation of the value to the individual within an hierarchical setting using the proposed model.}
        \label{vi_l}
    \end{figure}

\begin{figure}[tpb]
       \resizebox{\hsize}{!}{$\displaystyle{
          \centering{
                     \begin{tikzpicture}
                            \begin{axis}[
                                         title={Productive percentage by the value to the individual},
                                         axis lines = left,
                                         xlabel = {$v_i$ [\$]},
                                         ylabel = {$Productivity$ [\%]},
                                         legend pos=outer north east,
                                         ]
                              \addplot [
                                         domain=0:10, 
                                         samples=20, 
                                         color=red,
                                         mark=triangle, 
                                         mark size=2,
                                       ]
                                        {1 - e^(-2*x)};
                                      \addlegendentry{\text{\it{Expected state}}}
                              \addplot [
                                        domain=-0:10, 
                                        samples=20, 
                                        color=blue,
                                        mark=*,
                                        name path=A,
                                       ]
                                        {1 - e^(-x) };
                                       \addlegendentry{\text{\it{Typical state}}}
                                       
                              \addplot [
                                        domain=-0:10, 
                                        samples=20, 
                                        color=orange,
                                        mark=+, 
                                        name path=B,
                                       ]
                                        {1};
                                       \addlegendentry{\text{\it{Ideal state}}}                                 
                                                                 
                  \addplot+[green, fill opacity=0.2] fill between[of=A and B,soft clip={domain=0:10}];
                            \end{axis}
                     \end{tikzpicture}
                     }
                    }
             $}
          \caption{Estimation of the productivity as a function to the value to the individual using the proposed model.}
          \label{prod_vi}
     \end{figure}
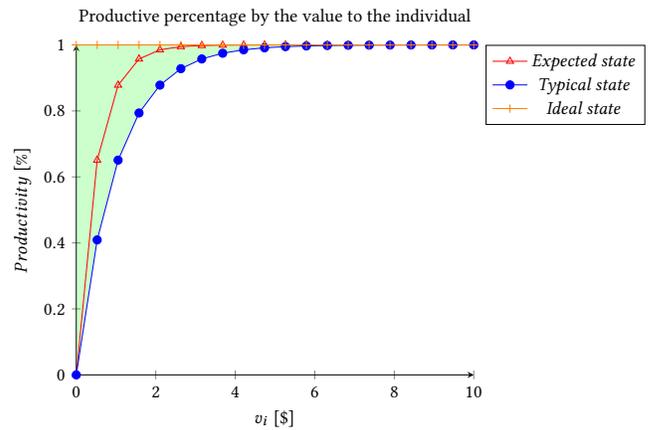

 Figure \ref{label_Flows} shows a schematic diagram of the leading entities and attributes of the proposed DO model; where DL stands for Distributed Ledgers. 
 Moreover, figure \ref{vi_l} and \ref{prod_vi} show the output for the value to the individual and productivity, respectively. The blue line shows the output that most of the organisations present; the yellow line indicates the ideal curve and the orange one displays the expected curves of the model. Finally, the green area exhibits an acceptable range for these variables after the implementation of the RO.
    
%
\section{Numerical example}
\begin{figure}[tpb]
            \centering{\includegraphics[scale=0.4]{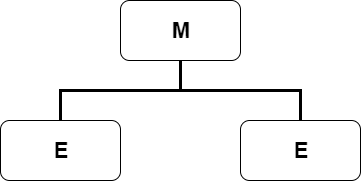}}
            \caption{Example's firm structure.}
            \label{label_ORG_EXAM}
        \end{figure}

 This passage demonstrates the proposed model. Considering the equation \ref{COASE_OLSON_EQU2} and the values $j=5$, $l=2$, $r=30\%$, $S=100$ £, $C=75$ £, $\alpha = 0.6$, and the firm's structure shown in the Fig.\ref{label_ORG_EXAM}.    
 With $w_{r_{1,3}} = 5$, $w_{r_{1,4}} = 4$, and $w_{r_{1,5}} = 3$. Where the manager, M, has the index 3 and the employees, E, 4 and 5. Then we have the following: $(S-C) = 25$, $\beta_{V_{0,0}} = 0.1$, $\beta_{V_{0,1}} = 0.2$, $\beta_{V_{1,3}} = 0.2221$, $\beta_{V_{1,4}} = 0.2389$, $\beta_{V_{1,5}} = 0.2389$,  $\tau_{1,{4}}$ = 1,  $\tau_{1,{5}}$ = 1, $\tau_{1,{1_4}}$ = 2, and  $\tau_{1,{1_5}}$ = 2. Thus, the value for  the individuals are: $V_{0,0} = 2.5$, $V_{0,1} = 5$, $V_{1,3} = 18.0528$, $V_{1,4} = 15.9736$, and $V_{1,5} = 13.4736$.
 
 As for the $TTC$, this value changes at every business cycle. Thus, we recommend continuously monitoring the different factors that affect this variable.

 \section{Conclusions and further research}
 In this article, we have introduced a new digital infrastructure: the digital cooperative. This Decentralised Organisation (DO) aims for balancing the distribution of the rewards and values to the individuals of the group and minimising the transaction costs with the support of smart contracts. We described the operational conditions of DO. We also analysed the introduction of a reward oracle (RO), within the context of a DO abstract model, that can induce the emergence of a new type of organisation.
 
 The notions of "value to the individual" and "productivity" helped us to explain the effects of this transformation. It is relevant to mention that RO is an abstract model which links plausible conditions for a digital infrastructure supporting a DO and the emerging microeconomic patterns. That is, RO aims to compute optimal conditions for an organisation to operate under a distributed and group-efficient environment.
 
 Further research should introduce the concept of uncertainty to the proposed equations in order to mimic market conditions. In other words, the integration of other mathematical tools such as stochastic processes or mean-field game theory. Moreover, homogeneous agent simulations will facilitate the identification of more specific qualification of microeconomic properties for distributed organisations. 
 
 Finally, decentralised economies and the economic impact of automation are phenomena which are still under development\cite{DeFilippi2018}. Their full understanding requires the creation of end-to-end economic models that embed some of the properties entailed by these digital infrastructures.

\bibliographystyle{ACM-Reference-Format}
\bibliography{bibliography} 


\begin{thebibliography}{17}


\ifx \showCODEN    \undefined \def \showCODEN     #1{\unskip}     \fi
\ifx \showDOI      \undefined \def \showDOI       #1{#1}\fi
\ifx \showISBNx    \undefined \def \showISBNx     #1{\unskip}     \fi
\ifx \showISBNxiii \undefined \def \showISBNxiii  #1{\unskip}     \fi
\ifx \showISSN     \undefined \def \showISSN      #1{\unskip}     \fi
\ifx \showLCCN     \undefined \def \showLCCN      #1{\unskip}     \fi
\ifx \shownote     \undefined \def \shownote      #1{#1}          \fi
\ifx \showarticletitle \undefined \def \showarticletitle #1{#1}   \fi
\ifx \showURL      \undefined \def \showURL       {\relax}        \fi
\providecommand\bibfield[2]{#2}
\providecommand\bibinfo[2]{#2}
\providecommand\natexlab[1]{#1}
\providecommand\showeprint[2][]{arXiv:#2}

\bibitem[\protect\citeauthoryear{Atrill and McLaney}{Atrill and
  McLaney}{2017}]%
        {Atrill2017}
\bibfield{author}{\bibinfo{person}{Peter Atrill} {and} \bibinfo{person}{Eddie
  McLaney}.} \bibinfo{year}{2017}\natexlab{}.
\newblock \bibinfo{booktitle}{\emph{{Accounting and Finance for
  Non-Specialists}} (\bibinfo{edition}{10} ed.)}.
\newblock \bibinfo{publisher}{Pearson Education Limited}. 617 pages.
\newblock
\showISBNx{9781292135601}


\bibitem[\protect\citeauthoryear{BACKFEED}{BACKFEED}{2016}]%
        {BACKFEED}
\bibfield{author}{\bibinfo{person}{BACKFEED}.} \bibinfo{year}{2016}\natexlab{}.
\newblock \bibinfo{title}{{BACKFEED: A Social Operating System for
  Decentralized Organizations}}.
\newblock
\newblock
\urldef\tempurl%
\url{http://backfeed.cc}
\showURL{%
\tempurl}


\bibitem[\protect\citeauthoryear{Buchanan}{Buchanan}{1990}]%
        {Buchanan1990}
\bibfield{author}{\bibinfo{person}{James~M. Buchanan}.}
  \bibinfo{year}{1990}\natexlab{}.
\newblock \showarticletitle{{The domain of constitutional economics}}.
\newblock \bibinfo{journal}{\emph{Constitutional Political Economy}}
  \bibinfo{volume}{1}, \bibinfo{number}{1} (\bibinfo{date}{dec}
  \bibinfo{year}{1990}), \bibinfo{pages}{1--18}.
\newblock
\showISSN{1043-4062}
\urldef\tempurl%
\url{https://doi.org/10.1007/BF02393031}
\showDOI{\tempurl}


\bibitem[\protect\citeauthoryear{Coase}{Coase}{1937}]%
        {Coase1937}
\bibfield{author}{\bibinfo{person}{R.~H. Coase}.}
  \bibinfo{year}{1937}\natexlab{}.
\newblock \showarticletitle{{The Nature of the Firm}}.
\newblock \bibinfo{journal}{\emph{Economica}} \bibinfo{volume}{4},
  \bibinfo{number}{16} (\bibinfo{date}{nov} \bibinfo{year}{1937}),
  \bibinfo{pages}{386--405}.
\newblock
\showISSN{0013-0427}
\urldef\tempurl%
\url{https://doi.org/10.1111/j.1468-0335.1937.tb00002.x}
\showDOI{\tempurl}


\bibitem[\protect\citeauthoryear{Davidson, {De Filippi}, and Potts}{Davidson
  et~al\mbox{.}}{2016}]%
        {Davidson2016}
\bibfield{author}{\bibinfo{person}{Sinclair Davidson},
  \bibinfo{person}{Primavera {De Filippi}}, {and} \bibinfo{person}{Jason
  Potts}.} \bibinfo{year}{2016}\natexlab{}.
\newblock \showarticletitle{{Economics of Blockchain}}.
\newblock \bibinfo{journal}{\emph{SSRN Electronic Journal}}
  (\bibinfo{year}{2016}).
\newblock
\showISSN{1556-5068}
\urldef\tempurl%
\url{https://doi.org/10.2139/ssrn.2744751}
\showDOI{\tempurl}


\bibitem[\protect\citeauthoryear{De~Filippi}{De~Filippi}{2018}]%
        {DeFilippi2018}
\bibfield{author}{\bibinfo{person}{Primavera De~Filippi}.}
  \bibinfo{year}{2018}\natexlab{}.
\newblock \bibinfo{booktitle}{\emph{{Blockchain and the law : the rule of
  code}}}.
\newblock \bibinfo{publisher}{Harvard University Press"},
  \bibinfo{address}{Cambridge, Massachusetts}.
\newblock
\showISBNx{9780674985933}


\bibitem[\protect\citeauthoryear{Hayek}{Hayek}{1988}]%
        {Hayek1988}
\bibfield{author}{\bibinfo{person}{F.A. Hayek}.}
  \bibinfo{year}{1988}\natexlab{}.
\newblock \bibinfo{booktitle}{\emph{{The Fatal Conceit. The Errors of
  Socialism.}}}
\newblock \bibinfo{publisher}{Routledge}, \bibinfo{address}{London: Routledge}.
\newblock
\showISBNx{0-415-00820-4}


\bibitem[\protect\citeauthoryear{Kantarelis}{Kantarelis}{2017}]%
        {Kantarelis2017}
\bibfield{author}{\bibinfo{person}{Demetri Kantarelis}.}
  \bibinfo{year}{2017}\natexlab{}.
\newblock \bibinfo{booktitle}{\emph{{Theories of the Firm}}
  (\bibinfo{edition}{fifth} ed.)}.
\newblock \bibinfo{publisher}{Inderscience Enterprise Ltd},
  \bibinfo{address}{London, UK}. 392 pages.
\newblock
\showISBNx{0-907776-62-0}


\bibitem[\protect\citeauthoryear{Marsh and Onof}{Marsh and Onof}{2008}]%
        {Marsh2008}
\bibfield{author}{\bibinfo{person}{Leslie Marsh} {and}
  \bibinfo{person}{Christian Onof}.} \bibinfo{year}{2008}\natexlab{}.
\newblock \showarticletitle{{Stigmergic epistemology, stigmergic cognition}}.
\newblock \bibinfo{journal}{\emph{Cognitive Systems Research}}
  \bibinfo{volume}{9}, \bibinfo{number}{1-2} (\bibinfo{date}{mar}
  \bibinfo{year}{2008}), \bibinfo{pages}{136--149}.
\newblock
\showISSN{13890417}
\urldef\tempurl%
\url{https://doi.org/10.1016/j.cogsys.2007.06.009}
\showDOI{\tempurl}


\bibitem[\protect\citeauthoryear{Morgan}{Morgan}{2008}]%
        {Morgan2008}
\bibfield{author}{\bibinfo{person}{Mary~S. Morgan}.}
  \bibinfo{year}{2008}\natexlab{}.
\newblock \showarticletitle{{Models}}.
\newblock In \bibinfo{booktitle}{\emph{The New Palgrave Dictionary of
  Economics}}. \bibinfo{publisher}{Palgrave Macmillan UK},
  \bibinfo{address}{London}, \bibinfo{pages}{1--14}.
\newblock
\urldef\tempurl%
\url{https://doi.org/10.1057/978-1-349-95121-5_2171-1}
\showDOI{\tempurl}


\bibitem[\protect\citeauthoryear{Olson}{Olson}{1965}]%
        {Olson1965}
\bibfield{author}{\bibinfo{person}{Mancur Olson}.}
  \bibinfo{year}{1965}\natexlab{}.
\newblock \bibinfo{booktitle}{\emph{{The Logic of Collective Action: Public
  Goodsand the Theory of Groups.}}}
\newblock \bibinfo{publisher}{Harvard University Press},
  \bibinfo{address}{London}. 208 pages.
\newblock


\bibitem[\protect\citeauthoryear{Pazaitis, {De Filippi}, and Kostakis}{Pazaitis
  et~al\mbox{.}}{2017}]%
        {Pazaitis2017}
\bibfield{author}{\bibinfo{person}{Alex Pazaitis}, \bibinfo{person}{Primavera
  {De Filippi}}, {and} \bibinfo{person}{Vasilis Kostakis}.}
  \bibinfo{year}{2017}\natexlab{}.
\newblock \showarticletitle{{Blockchain and value systems in the sharing
  economy: The illustrative case of Backfeed}}.
\newblock \bibinfo{journal}{\emph{Technological Forecasting and Social Change}}
   \bibinfo{volume}{125} (\bibinfo{date}{dec} \bibinfo{year}{2017}),
  \bibinfo{pages}{105--115}.
\newblock
\showISSN{00401625}
\urldef\tempurl%
\url{https://doi.org/10.1016/j.techfore.2017.05.025}
\showDOI{\tempurl}


\bibitem[\protect\citeauthoryear{Sokolowsky}{Sokolowsky}{2015}]%
        {SOKOLOWSKY2015}
\bibfield{author}{\bibinfo{person}{Oren Sokolowsky}.}
  \bibinfo{year}{2015}\natexlab{}.
\newblock \bibinfo{title}{{La`Zooz}}.
\newblock
\newblock
\urldef\tempurl%
\url{http://lazooz.org/}
\showURL{%
\tempurl}


\bibitem[\protect\citeauthoryear{Suematsu}{Suematsu}{2014}]%
        {Suematsu2014}
\bibfield{author}{\bibinfo{person}{Chihiro Suematsu}.}
  \bibinfo{year}{2014}\natexlab{}.
\newblock \bibinfo{booktitle}{\emph{{Transaction Cost Management}}}.
\newblock \bibinfo{publisher}{Springer International Publishing},
  \bibinfo{address}{Cham}.
\newblock
\showISBNx{978-3-319-06888-6}
\urldef\tempurl%
\url{https://doi.org/10.1007/978-3-319-06889-3}
\showDOI{\tempurl}


\bibitem[\protect\citeauthoryear{Team}{Team}{2017}]%
        {Team2017}
\bibfield{author}{\bibinfo{person}{The~CORE Team}.}
  \bibinfo{year}{2017}\natexlab{}.
\newblock \bibinfo{booktitle}{\emph{{The Economy: Economics for a Changing
  World}} (\bibinfo{edition}{first} ed.)}.
\newblock \bibinfo{publisher}{OXFORD University Press}, \bibinfo{address}{UK}.
  1152 pages.
\newblock
\showISBNx{9780198810247}


\bibitem[\protect\citeauthoryear{{UK Government Chief Scientific Adviser}}{{UK
  Government Chief Scientific Adviser}}{2016}]%
        {UKGovernmentChiefScientificAdviser2016}
\bibfield{author}{\bibinfo{person}{{UK Government Chief Scientific Adviser}}.}
  \bibinfo{year}{2016}\natexlab{}.
\newblock \bibinfo{booktitle}{\emph{{Distributed Ledger Technology: beyond
  block chain}}}.
\newblock \bibinfo{type}{{T}echnical {R}eport}.
  \bibinfo{institution}{Government Office for Science},
  \bibinfo{address}{London, UK}. \bibinfo{pages}{88} pages.
\newblock
\urldef\tempurl%
\url{https://bit.ly/2lpNHhD}
\showURL{%
\tempurl}


\bibitem[\protect\citeauthoryear{Williamson}{Williamson}{1998}]%
        {Williamson1998}
\bibfield{author}{\bibinfo{person}{Oliver~E. Williamson}.}
  \bibinfo{year}{1998}\natexlab{}.
\newblock \showarticletitle{{Transaction Cost Economics: How It Works; Where It
  is Headed}}.
\newblock \bibinfo{journal}{\emph{De Economist}} \bibinfo{volume}{146},
  \bibinfo{number}{1} (\bibinfo{year}{1998}), \bibinfo{pages}{23--58}.
\newblock


\end{thebibliography}

\end{document}